\documentclass{aa}  

\usepackage{linenoaa}

\usepackage{graphicx}
\usepackage{txfonts}
\usepackage{booktabs}
\usepackage{titlesec}
\usepackage[dvipsnames]{xcolor}
\usepackage{subcaption}
\usepackage{pdflscape}

\usepackage{color}
\usepackage{natbib,twoopt}
\usepackage[hyphenbreaks]{breakurl}
\usepackage[breaklinks]{hyperref}      
\bibpunct{(}{)}{;}{a}{}{,}             
\definecolor{cobalt}{rgb}{0.06, 0.2, 0.65}
\hypersetup{
  colorlinks,
  citecolor=blue,
  linkcolor=blue,
  urlcolor=blue,
}
\makeatletter
  \newcommandtwoopt{\citeads}[3][][]{\href{http://adsabs.harvard.edu/abs/#3}%
    {\def\hyper@linkstart##1##2{}%
     \let\hyper@linkend\@empty\citealp[#1][#2]{#3}}}
  \newcommandtwoopt{\citepads}[3][][]{\href{http://adsabs.harvard.edu/abs/#3}%
    {\def\hyper@linkstart##1##2{}%
     \let\hyper@linkend\@empty\citep[#1][#2]{#3}}}
  \newcommandtwoopt{\citetads}[3][][]{\href{http://adsabs.harvard.edu/abs/#3}%
    {\def\hyper@linkstart##1##2{}%
     \let\hyper@linkend\@empty\citet[#1][#2]{#3}}}
  \newcommandtwoopt{\citeyearads}[3][][]%
    {\href{http://adsabs.harvard.edu/abs/#3}
    {\def\hyper@linkstart##1##2{}%
     \let\hyper@linkend\@empty\citeyear[#1][#2]{#3}}}
\makeatother

\titleformat{\paragraph}
{\normalfont\normalsize}{\theparagraph}{1em}{}
\titlespacing*{\paragraph}
{0pt}{3.25ex plus 1ex minus .2ex}{1.5ex plus .2ex}

\newcommand{\bse}{BSE\xspace}
\newcommand{\mocca}{MOCCA\xspace}
\newcommand{\fewbody}{FEWBODY\xspace}

\titlerunning{GW signals from DWDs inside GCs}
\authorrunning{L. Hellström et al.}

\begin{document}

   \title{Formation Channels of Gravitationally Resolvable Double White Dwarf Binaries Inside Globular Clusters}

   \author{
    L. Hellström$^{1}$,
    M. Giersz$^{1}$, A. Askar$^{1}$, A Hypki$^{1,2}$, Y. Zhao$^{3}$, Y. Lu$^{4,5}$,  S. Zhang$^{4,5}$, V. V\'{a}zquez-Aceves$^6$, G. Wiktorowicz$^{1}$
    }

   \institute{Nicolaus Copernicus Astronomical Center, Polish Academy of Sciences, ul. Bartycka 18, PL-00-716 Warsaw, Poland
   \\
              \email{hellstrom@camk.edu.pl}
             \and
   Faculty of Mathematics and Computer Science, A. Mickiewicz University, Uniwersytetu Pozna\'nskiego 4, 61-614 Pozna\'n, Poland
   \and
   School of physics and astronomy, Beijing Normal University, 19 Xinjiekouwai St, Beijing 100875, People’s Republic of China
   \and
    National Astronomical Observatories, Chinese Academy of Sciences, 20A Datun Road, Beijing 100101, People’s Republic of China
    \and 
     School of Astronomy and Space Sciences, University of Chinese Academy of Sciences, 19A Yuquan Road, Beijing 100049, People’s Republic of China
     \and
     Kavli Institute for Astronomy and Astrophysics at Peking University, 100871 Beijing, China
             }

   \date{Accepted XXX. Received YYY; in original form ZZZ}

  \abstract{Current gravitational wave detectors are sensitive to coalescing black holes and neutron stars. However, double white dwarfs (DWDs) have long been recognized as promising sources of gravitational waves, and upcoming detectors like LISA will be able to observe these systems in abundance. DWDs are expected to be the dominant gravitational wave (GW) sources in parts of the LISA frequency range, making it crucial to understand their formation for future detections. The Milky Way contains many white dwarfs (WDs) in both the field and star clusters, promising a rich population of DWDs for LISA. However, the large number of sources may make it difficult to resolve individual binaries, and DWDs in the field and clusters often have similar properties, complicating the identification of their origins from GW signals alone. In this work, we focus on eccentric and tight DWDs, which cannot form in the field, but require dynamical interactions in dense clusters to increase their eccentricity after circularization through mass transfer phases and common-envelope evolution during binary evolution. These binaries may also form in three- and four-body dynamical interactions where a DWD binary may directly form with high eccentricity and low separation. Our results show that we should expect eccentric and tight DWDs in clusters that can provide meaningful GW signal, however, the number is low; with an upper limit of 10-15 in the MW. These can be used to independently obtain distances of their host cluster.}


   \maketitle

%

\section{Introduction}
Current gravitational wave (GW) detectors are primarily sensitive to the frequency range of merging black holes (BHs) and neutron stars (NSs), and the majority of GW discoveries to date have involved these systems. However, double white dwarf (DWD) binaries are also important sources of gravitational waves. Their signals occur at lower frequencies, outside the sensitivity range of current ground-based detectors, which is why no GW detections from DWDs have yet been made. The planned launch of space-based observatories such as LISA \citep{LISA_2017, lisa_2024}, Taiji \citep{Ruan2018}, and TianQin \citep{TianQin_2016, TianQin_2025} will make it possible to detect GWs from DWDs for the first time.

DWDs are multi-messenger sources and can today be observed in optical, ultraviolet, x-ray and potentially radio waves \citep{deRuiter2025}. With the introduction of space-based detectors, it will also be possible to observe them through GWs. As WDs are very common objects, with 97\% of stars in the Milky Way estimated to become WDs \citep{fontaine2001}, the probability to detect such a signal increases. DWDs can be used for distance measuring and GW observations are another possible way to determine the distance to a DWD. For DWDs inside clusters we may not able to individually observe them, especially if the orbit is very tight and the two components cannot be individually electromagnetically resolved. Due to this, GWs might be an effective way of discovering DWDs inside clusters. The population of GW observable DWDs is expected to be large \citep{lamberts2019L}, however, the number of WDs introduces another challenge since too many objects may lead to unresolved signals \citep{ruiter2010, nissanke2012}. This background signal is something that is expected to be found with LISA and will cause many DWDs to be unresolvable. However, in some cases the orbit of a DWD can be very tight and eccentric and thus the signal will be stronger and potentially resolvable. 

There is also a foreground signal from close-by tight field DWDs \citep{ruiter2010,korol2022, thiele2023, toubiana2024} that is expected to dominate certain frequency ranges. For circular binaries it would not be possible to distinguish a signal from a field and a cluster binary. However, eccentric and tight DWDs cannot form solely through binary evolution due to circularization during the evolution of the binary. Although, triple systems with an inner DWD binary exists in the field \citep{perpiny2019, ArosBunster2025} and the addition of a third component may increase the eccentricity of the inner binary \citep{naoz2013, antognini2014, liu2015}. Combined with a small separation this may produce an eccentric DWD that can be observed with LISA \citep{rajamuthukumar2025}. We will focus on eccentric DWDs in this paper which is a binary property that can be used to differentiate between binaries inside clusters and in the field and we will discuss more about the triple channel later in the paper.

White dwarfs and double white dwarfs are expected to be common in globular clusters (GCs). This is because GCs are very old, allowing stars with initially low mass to have enough time to evolve into WDs. In contrast, in younger regions, such as the galactic disk or open clusters (OCs), low-mass stars are still in the main sequence phase and haven't yet evolved into WDs. In clusters we can expect to find DWDs that have hardened in dynamical interactions. However, GCs are, on average, farther away from us than OCs or near-by field stars. Since the strength of GWs are reduced over distance, closer distances are preferred when observing DWD in GWs. The age of a cluster also affects the objects inside it, in an older cluster there is more time for lower mass stars to evolve into WDs, however, at later times tight WDs might have merged or could have escaped the cluster leading to more WDs with smaller masses. GCs are typically very old with no ongoing star formation, on the other hand, OCs are on average younger and might contain different types of binaries due to this. Both kinds of clusters have the possibility to contain tight, low period DWDs, however in this paper we will focus on GCs. OCs are generally closer to us and thus the SNR is higher, however, the mass of GCs are higher to compensate for the increased distance with more potential eccentric sources. In addition to this, there has been shown that massive WDs dominate the inner region of core collapsed GCs \citep{kremer2021}, allowing massive WDs to form tight binaries which may then be involved in dynamical interactions, further hardening and/or increasing the eccentricity of the DWD. 

There is a range in the frequency spectrum ($\sim0.1$ mHz to $\sim10$ mHz) \citep{nelemans2001, korol2017} that is likely dominated by DWDs. Although, this frequency range is inaccessible to ground-bases GW detectors, the future space based detector such as LISA, Taiji, and TianQin will operate in a lower frequency range and will be sensitive to weaker signals, allowing us to detect DWDs not only during the merger but also during their inspiral phase, where the frequencies fall within their sensitivity peak. In fact, the continuous signal from a very tight DWD is much more likely to be detected due to the large number of such binaries in the MW.

The idea of observing DWDs that are inside dense environments with GWs have previously been done, however with limitations. The dynamical formation channel has been explored \citep[e.g.][]{willems2007} and studies on clusters where dynamical interactions were neglected has been looked into \citep[e.g.][e.g.]{vanZeist2024}. In this paper we analyze the formation channels of DWDs in GCs. As the properties of DWDs formed in GCs are unique to dense environments, we explore if these systems can be resolved and differentiated from the Galactic DWDs population via their GW signal. The main differences between earlier studies is that we use models containing multiple stellar populations, affecting the initial conditions of these clusters, we consider both primordial and dynamically formed binaries and we do not neglect dynamical interactions.

The aim of the paper is to explore the idea of finding binaries that are distinct from the galaxy population of DWDs, are unique to dense environments and are able to be observed in GWs. We will present a general analysis where we discover the possibility to find eccentric DWDs that are resolvable in GWs and their formation channels. In a future paper we will analyze the GW signals of these binaries in detail in order to obtain the properties of the binaries as well as distance estimations and sky localizations. The paper is structured as follows. In section \ref{sec:numSim} we present the numerical simulations and initial parameters used. Section \ref{sec:results} contains our results and in section \ref{sec:discussion} we discuss these results and present our conclusions.


\section{Models and initial conditions}
\label{sec:numSim}
\subsection{The \mocca code}
We use the MOnte Carlo Cluster simulAtor, \mocca\footnote{\url{http://moccacode.net}} \citep{MOCCA1,MOCCA2, hypki2025, giersz2025}, and more specifically the simulations from MOCCA-SURVEY Database III \citep{hypki2025, giersz2025}, which is an upgrade over the MOCCA-SURVEY Database II \citep{hypki2022} and MOCCA-SURVEY Database I \citep{moccaSurvey1}. \mocca is a GC simulator built on a Monte Carlo approach. The newest \mocca version allows the usage of multiple stellar populations with different properties, provides large control over initial cluster conditions, and allows quick evolution of large clusters with more than one million initial members. \fewbody \citep{freqeau2004, freqeau2007} is used to solve strong dynamical interactions between binaries and stars as well as between binaries and binaries. 
   
An important aspect of the \mocca~code for our purposes is how stellar and binary evolution is treated, which is done with the modified \bse~code \citep{Hurley2000,Hurley2002}.

The \bse~code consists of a set of algorithms describing single star evolution, from zero-age main-sequence stars to later stages of stellar evolution, and binary evolution, taking into account angular momentum loss mechanisms, different modes of mass transfer, and tidal interaction.

\bse~has been widely used to investigate different astrophysical objects and is characterized by its generally high level of accuracy in the analytic fitting formulae on which it is based.

Since the publication of \bse, there have been several upgrades to the code \citep[see, e.g.,][for more details]{kamlah2022}. The main upgrades that are relevant for this study are the inclusion of improved wind prescriptions \citep{belczynski2016}, improved treatment of the common-envelope phase by adopting a $\alpha \lambda$ formalism \citep{giacobbo2018}, inclusion of a proper prescription for cataclysmic variable evolution \citep{belloni2018b,belloni2019} and improved remnant mass prescriptions for BHs and NSs \citep{banerjee2020}. 

\subsection{Initial conditions}
We use a total of 185 different GC models that survive longer than 9 Gyrs. A full description of the different cluster initial conditions can be found in \citet{hypki2025}. An important aspect of these cluster models is that they contain multiple stellar populations (MSPs) but no time delay in the introduction of the second population, as such we introduce both populations at $t=0$. This leads to an initial cluster structure composed of two distinct populations: one that is tidally filling or slightly tidally underfilling, less dense, and characterized by a higher maximum initial stellar mass; and another that is tidally underfilling, more compact, and composed of stars with lower maximum initial masses due to which there are no supernovae in this population. Additionally, because the pristine population is tidally filling, there is a large number of early escapers from the first population where approximately 30-40\% of mass is removed in the first few Myrs. When using the more traditional approach of one population the cluster is more dense and tidally underfilling and thus have less escapers than models with multiple populations.

The most important parameters for the GC models are the following:
\begin{itemize}
    \item A range of initial cluster sizes, from 550 thousand to 2.2 million stellar objects, modeled with two stellar populations
    \item Both tidally filling and tidally underfilling initial cluster configurations;
    \item Two initial binary fractions: 10\% and 95\%
    \item A Kroupa mass function between 0.08 and 150 M$_{\odot}$ (population 1) and 0.08 and 20 M$_{\odot}$ (population 2)
    \item Different Galactocentric distances between 1 and 4 kpc
    \item Different properties of multiple stellar populations (for details see \cite{hypki2025})
\end{itemize}

The initial properties of binaries are important for our study since it strongly affects the final outcome of binaries. Previous studies \citep{hypki2025} show that these models are good approximations for the properties of MW clusters.

In this work, we also analyzed 70 low-$N$ star cluster models with a single population with initial properties distinct from previous large-$N$ MOCCA surveys. This was done in order to explore if eccentric DWD binaries can potentially be found in old open OCs as well as dissolved GCs, however, our analysis on these models is brief since the focus on this paper is on Galactic GCs that have survived. The initial number of objects in these low-$N$ models ranges from $3.4 \times 10^4$ to $1.2 \times 10^5$, with a median value of $7.5 \times 10^4$. The corresponding initial cluster masses span $3.97 \times 10^4\,M_\odot$ to $1.19 \times 10^5\,M_\odot$ (median $6.91 \times 10^4\,M_\odot$). The initial half-mass radii range from 1.0 to 5.0\,pc (median 3.0\,pc), and core radii from 0.29 to 1.32\,pc (median 0.72\,pc). The initial central densities span $7.8 \times 10^2\,M_\odot\,\mathrm{pc}^{-3}$ to $1.6 \times 10^5\,M_\odot\,\mathrm{pc}^{-3}$ (median $8.7 \times 10^3\,M_\odot\,\mathrm{pc}^{-3}$). Initial tidal radii range from 10.4 to 87.5\,pc (median 59.0\,pc), corresponding to Galactocentric radii ranging from 1 to 16\,kpc (median 10.0\,kpc). The initial binary fraction in these models ranges from 10\% to 95\%, including 17 models with 40\%, 34 with 50\%, and 16 with 95\%. All models in this subset are metal poor, with $Z = 6.5 \times 10^{-5}$ (51 models) or $Z = 1.0 \times 10^{-4}$ (19 models), corresponding to $\lesssim 0.5\%$ of the solar value. The initial conditions for these models were adopted to represent old, metal-poor, and low-mass star clusters, motivated by the inferred properties for the progenitor star cluster of the ED-2 stream associated with the recently detected Gaia BH-3 \citep{2024A&A...686L...2G,2024A&A...687L...3B}.

\subsection{Gravitational wave simulations}
We use LEGWORK \citep{LEGWORK_apjs} for all GW calculations. LEGWORK is a python package that calculates analytical GW signals for binaries that could potentially be observed by LISA. LEGWORK calculates the signal-to-noise ratio $\rho$ as:
\begin{equation}
\langle \rho \rangle^2_{(\theta,\phi,\psi),(\iota,\beta)} = \sum_{n = 1}^{\infty} \langle \rho_n \rangle^2_{(\theta,\phi,\psi),(\iota,\beta)} = \sum_{n = 1}^{\infty} \int_0^{\infty} d f_n \frac{h_{c, n}^2}{f_n^2 S_{\rm n}(f_n)},
\label{eq:snr_general}
\end{equation}
\noindent where $\theta$ and $\phi$ define the position of the source on the sky, $\psi$ define the orientation of the source and $\iota$ and $\beta$ define the direction from the source to the detector. $f_n = n \cdot f_{orb}$, $n$ is the dominant harmonic, $f_{\rm{orb}}$ is the orbital frequency. $S_n$ is the effective LISA noise power spectral density \citep{robson2019} and $h_{c,n}$ is the characteristic strain amplitude, given by
\begin{equation}
h_{c,n}^2(\mathcal{M}_c, D_L, f_{\rm orb}, e) = \frac{2^{5/3}}{3 \pi^{4/3}} \frac{(G \mathcal{M}_c)^{5/3}}{c^3 D_L^2} \frac{1}{f_{\rm orb}^{1/3}} \frac{g(n, e)}{n F(e)},
\label{eq:char_strain}
\end{equation}
\noindent where $G$ is the gravitational constant, $\mathcal{M}_c$ is the chirp mass, $D_L$ is the luminosity distance, $g(n,e)$ is the relative GW power radiated into the $n^{th}$ harmonic from \cite{peters1963}:
\begin{equation}
\begin{aligned}
g(n, e)=\frac{n^4}{32} & \left\{\left[J_{n-2}(n e)-2 e J_{n-1}(n e)+\frac{2}{n} J_n(n e)\right.\right. \\
& \left.+2 e J_{n+1}(n e)-J_{n+2}(n e)\right]^2 \\
& +\left(1-e^2\right)\left[J_{n-2}(n e)-2 J_n(n e)\right. \\
& \left.+J_{n+2}(n e)\right]^2 \\
& \left.+\frac{4}{3 n^2}\left[J_n(n e)\right]^2\right\}
\end{aligned}
\end{equation}
\noindent where $J_n(v)$ is the Bessel function of the first kind \citep{watson1966bessel} and $F(e)$ is the enhancement factor from the same paper:
\begin{equation}
F(e) \equiv \sum_{n=1}^{\infty} g(n, e)=\frac{1+\frac{73}{24} e^2+\frac{37}{96} e^4}{\left(1-e^2\right)^{7 / 2}}.
\end{equation}

\noindent For eccentric binaries, the strain is summed up for a number of harmonics, where the number of harmonics needed depends on the eccentricity of the binary. This is a quick overview of the process of obtaining the SNR. For all details see \cite{LEGWORK_apjs}. For our study we position our clusters at random localizations in the sky at 2 kpc from the observer and set the observation time to 5 years.


\section{GW observations}
\label{sec:results}

\begin{figure*}[htbp]
    \centering
    \begin{subfigure}[b]{0.5\textwidth}
      \centering
      \includegraphics[width=.95\linewidth]{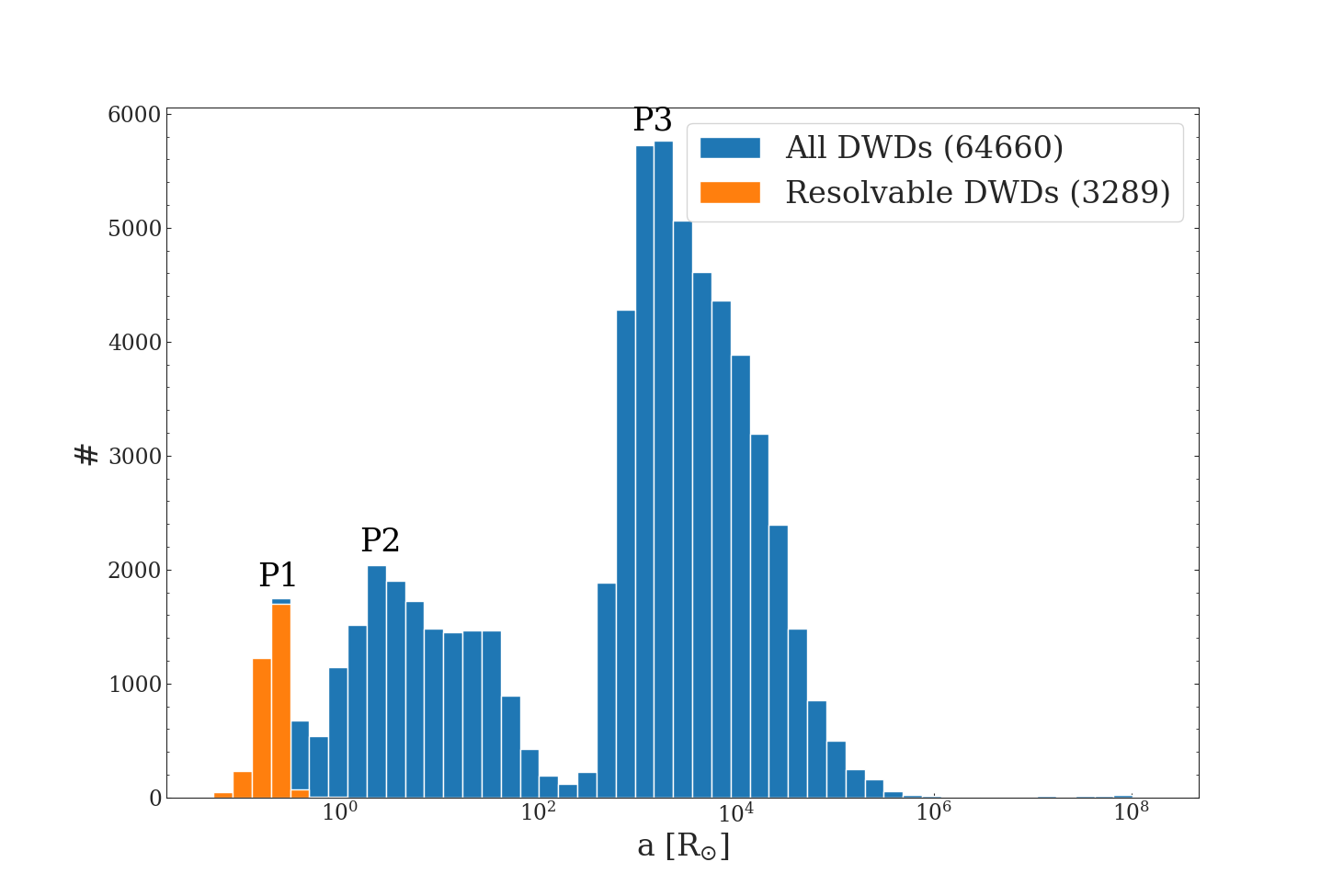}
      \caption{}
    \end{subfigure}%
    \begin{subfigure}[b]{0.5\textwidth}
      \centering
      \includegraphics[width=.95\linewidth]{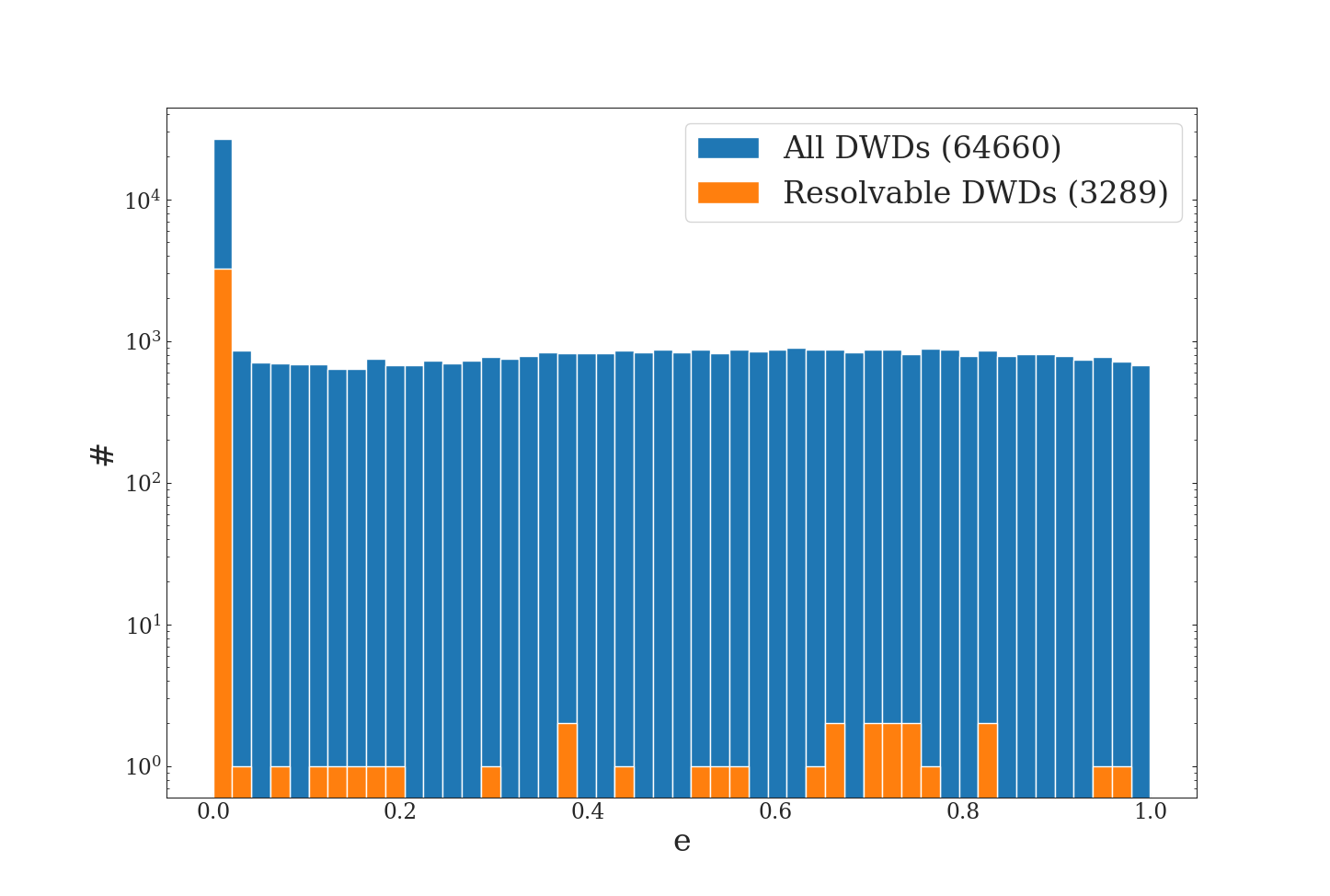}
      \caption{}
    \end{subfigure}

    \begin{subfigure}[b]{0.5\textwidth}
      \centering
      \includegraphics[width=.95\linewidth]{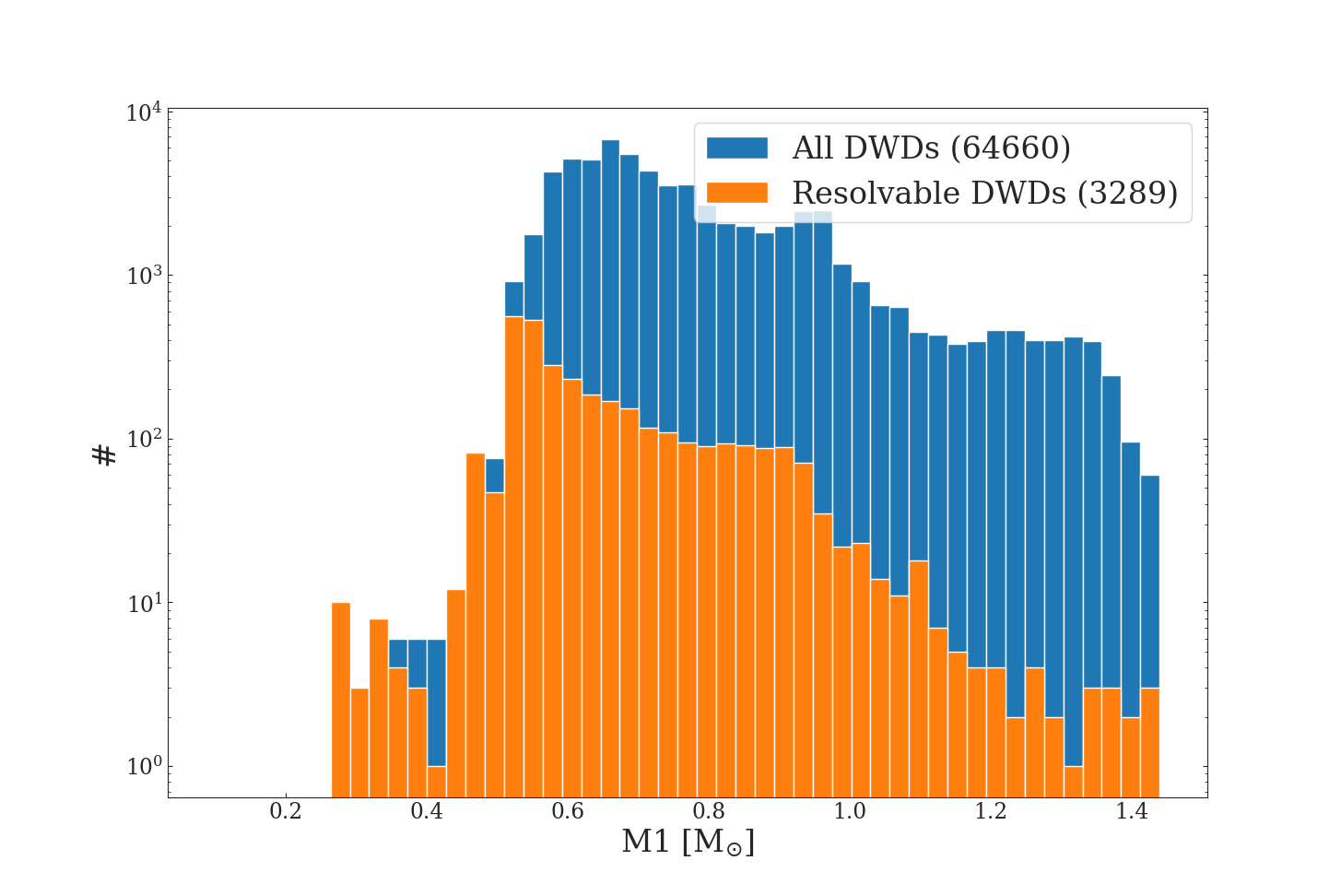}
      \caption{}
    \end{subfigure}%
    \begin{subfigure}[b]{0.5\textwidth}
      \centering
      \includegraphics[width=.95\linewidth]{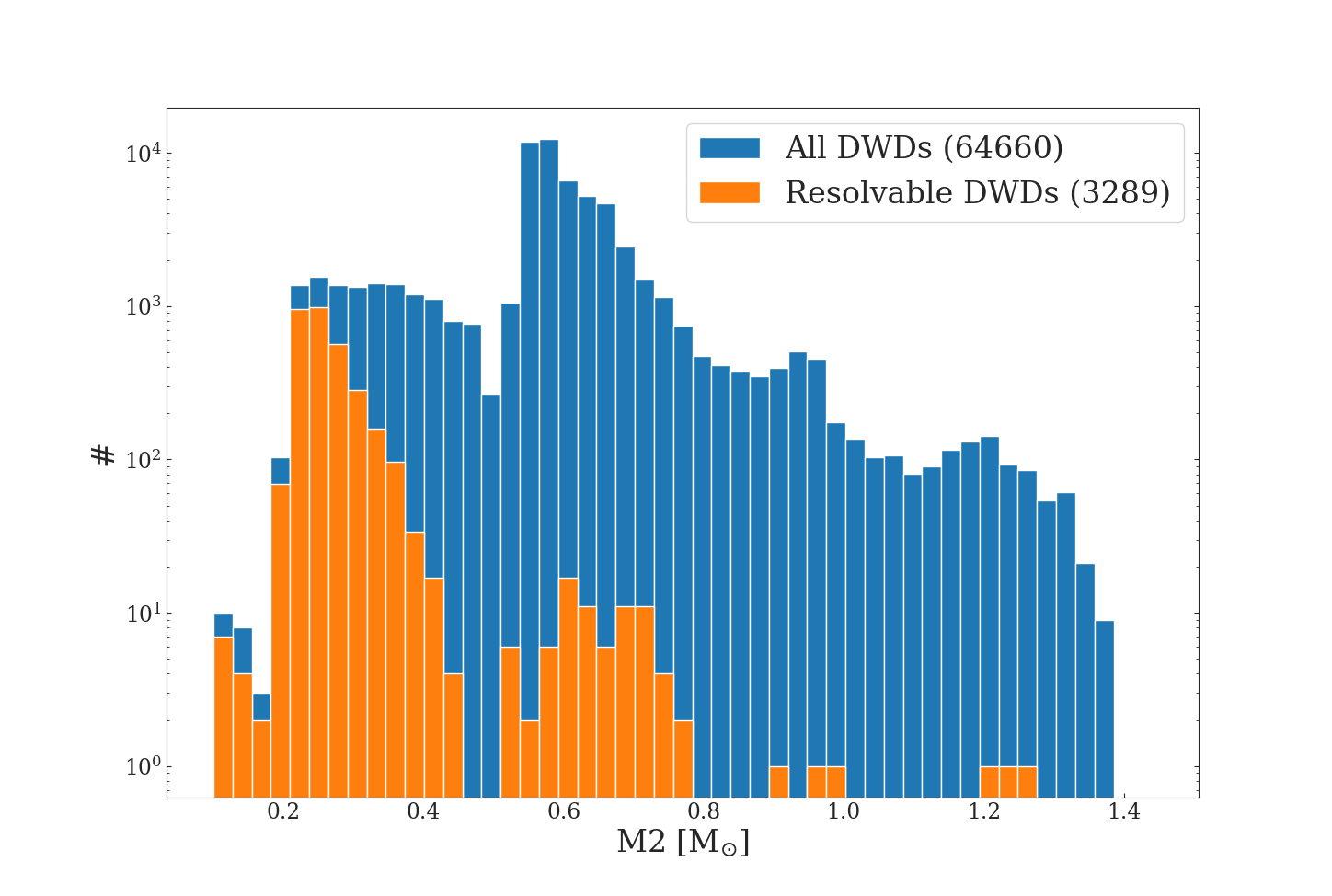}
      \caption{}
    \end{subfigure}
    \begin{subfigure}[b]{0.5\textwidth}
      \centering
      \includegraphics[width=.95\linewidth]{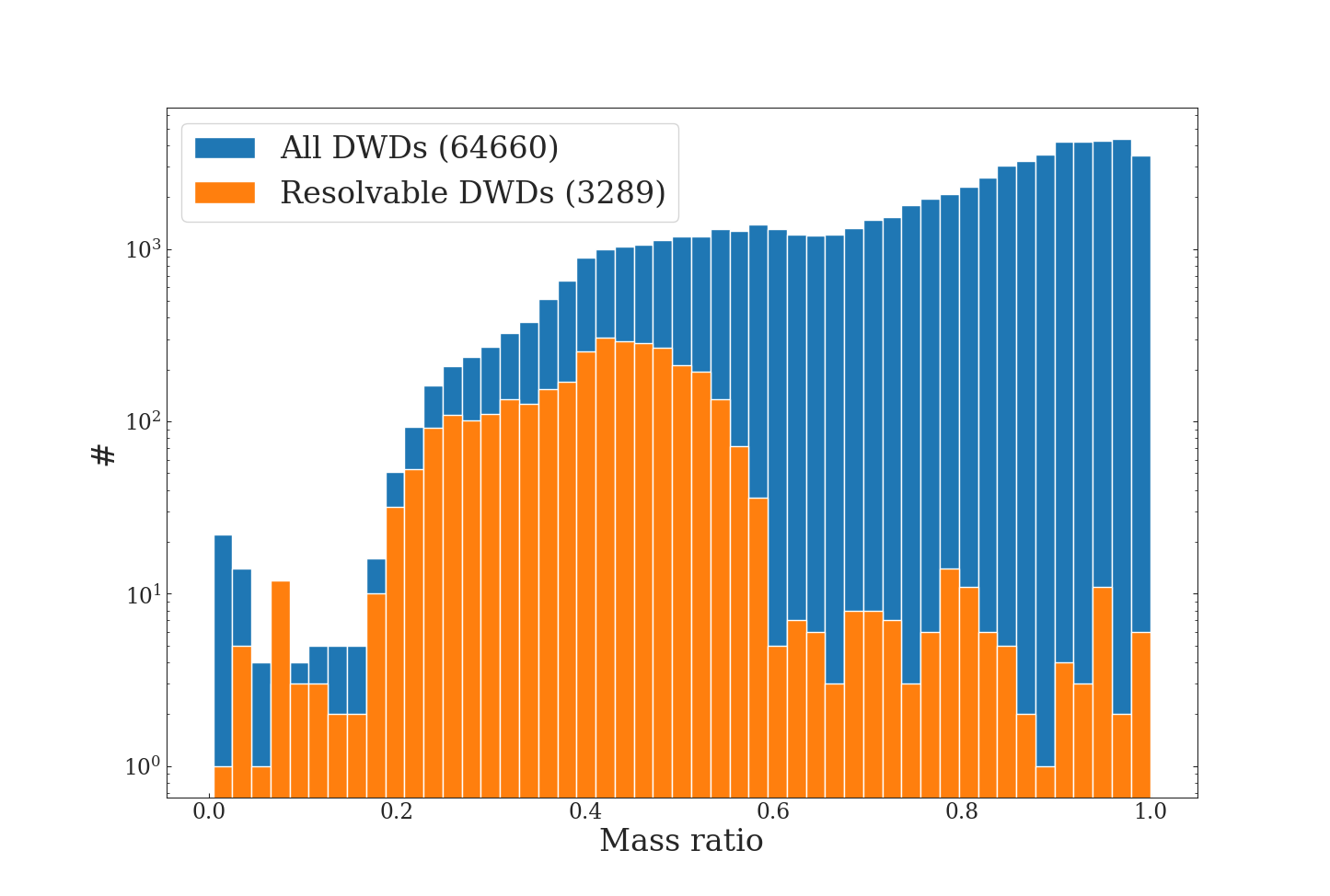}
      \caption{}
    \end{subfigure}%
    \caption{Histograms of the binary property distributions. The blue bars represents the whole dataset (64660 binaries) while the orange bars represents the resolvable binaries (3289 binaries). Panel a) shows the semi-major axis, panel b) shows the eccentricity, panels c) and d) shows the primary and secondary mass respectively and panel e) shows the mass ratio $\left(\frac{M2}{M1}\right)$.}
    \label{fig:binaryProperties_all}
\end{figure*}

Since we are interested in globular clusters that could potentially exist today we extract binaries from the clusters above 9 Gyrs up to 14 Gyrs. This gives a dataset which contains a total of 64660 DWDs. We are interested in resolvable binaries that are distinguishable from the GW background. We put all cluster models at 2 kpcs from the observer and observe them for 5 years. We limit ourselves to binaries with a LISA signal-to-noise, SNR, of above 10  \citep{ruiter2010, cornish2017} and end up with 3289 resolvable binaries from 185 cluster models in the orbital frequency range $10^{-5}$ to $10^{-2}$ Hz. 

The binary parameters of these binaries at the time when they are resolvable can be seen in Fig.\ref{fig:binaryProperties_all} where all 64660 binaries are shown in blue and the resolvable subset is shown in orange. Panel a) shows the semi-major axis, we can see that only the tightest binaries are resolvable. In addition to this we can see three groups in the big dataset as three distinct peaks, marked as P1, P2 and P3. By looking into these peaks we find that P3 consists of initially wide binaries where the components have evolved in isolation due to the large distances between them. P2 has interactions between the two binary components but the mass transfer is lower than for P1 and in most cases we only see one common-envelope phase. In P1 we have strong mass transfer during the evolution of the components and two common-envelope phases which tightens and circularizes the binaries. This peak may also contain DWDs that were directly formed with low semi-major axis in dynamical interactions and is the peak where we find all resolvable DWDs. 

In panel b) we show the eccentricity distribution; for all binaries the distribution is fairly even with a large bias towards 0 caused by circularization during the binary evolution. The eccentricities of binaries are initially thermally distributed and we can see that our distribution is not thermal. We do not see this for DWDs due to circularization that reduces the eccentricities of binaries. For resolvable binaries we have a very large bias towards 0 since these binaries are very tight and have gone through common-envelope phases. There are a few resolvable binaries with larger eccentricities that will be the focus of this paper. 

Panel c) shows the mass of the primary binary component, we can see that the distribution of masses for the resolvable DWDs are similar to the distribution of all DWDs with a peak around 0.7 M$_{\odot}$, a sharp short tail towards smaller masses and a longer tail towards higher masses up to the Chandrasekhar limit. For the secondary mass, shown in panel d),  there is a clear bias towards lower masses for the resolvable binaries that we do not see for all binaries. For all binaries we have more evenly distributed masses with a slight bias towards lower masses and a peak at around 0.6 M$_{\odot}$. This may be due to mass transfer onto the primary component during the binary evolution. The primary component evolves first into a WD and when the secondary expands into the giant phase it's likely that there will be mass transfer and a second common envelope phase due to the very small separations. This difference is also seen in panel e) where we show the mass ratio $\left(\frac{M2}{M1}\right)$. For all DWDs there is a bias towards higher mass ratio while for resolvable binaries the mass ratio drops sharply at around 0.6 and there are only a few resolvable binaries with mass ratio > 0.6. There is a peak for both groups at very low mass ratios (< 0.1), which is associated with binaries that have undergone substantial total mass transfer over time such as cataclysmic variables \citep{deloye2003, sion2023, wong2021}.

\begin{figure*}[htbp]
    \centering
    \begin{subfigure}[b]{0.5\textwidth}
      \centering
      \includegraphics[width=.95\linewidth]{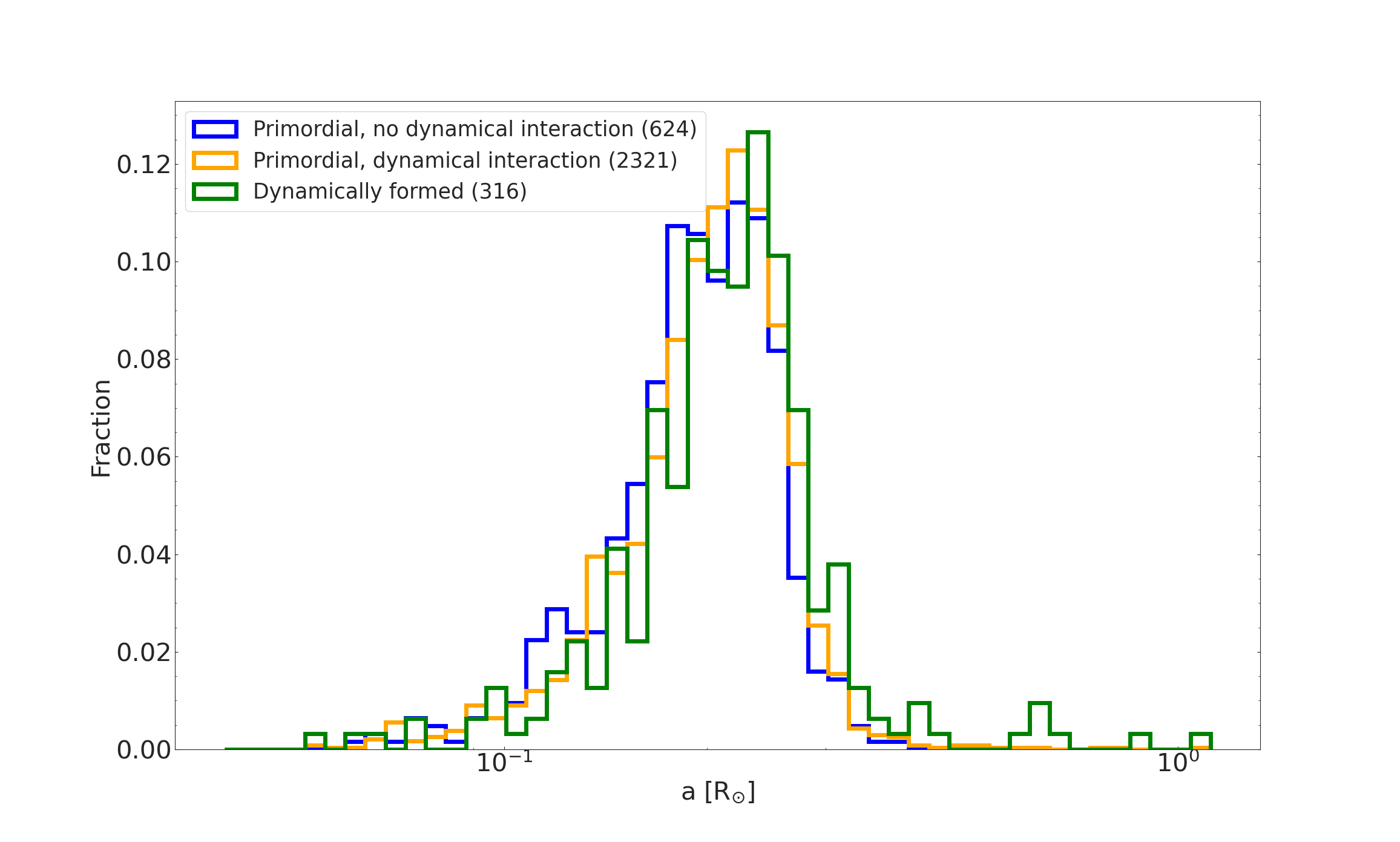}
      \caption{}
    \end{subfigure}%
    \begin{subfigure}[b]{0.5\textwidth}
      \centering
      \includegraphics[width=.95\linewidth]{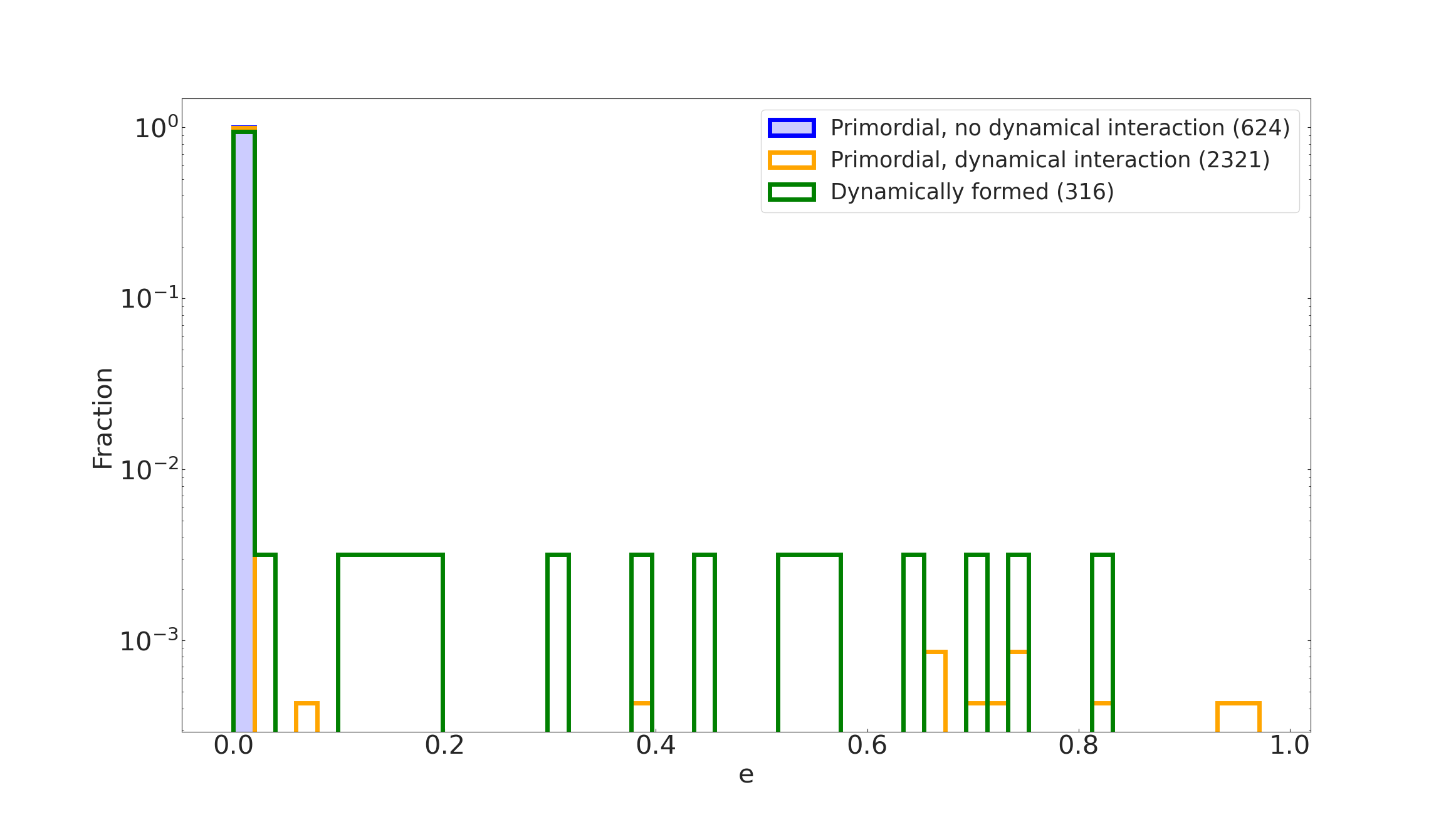}
      \caption{}
    \end{subfigure}

    \begin{subfigure}[b]{0.5\textwidth}
      \centering
      \includegraphics[width=.95\linewidth]{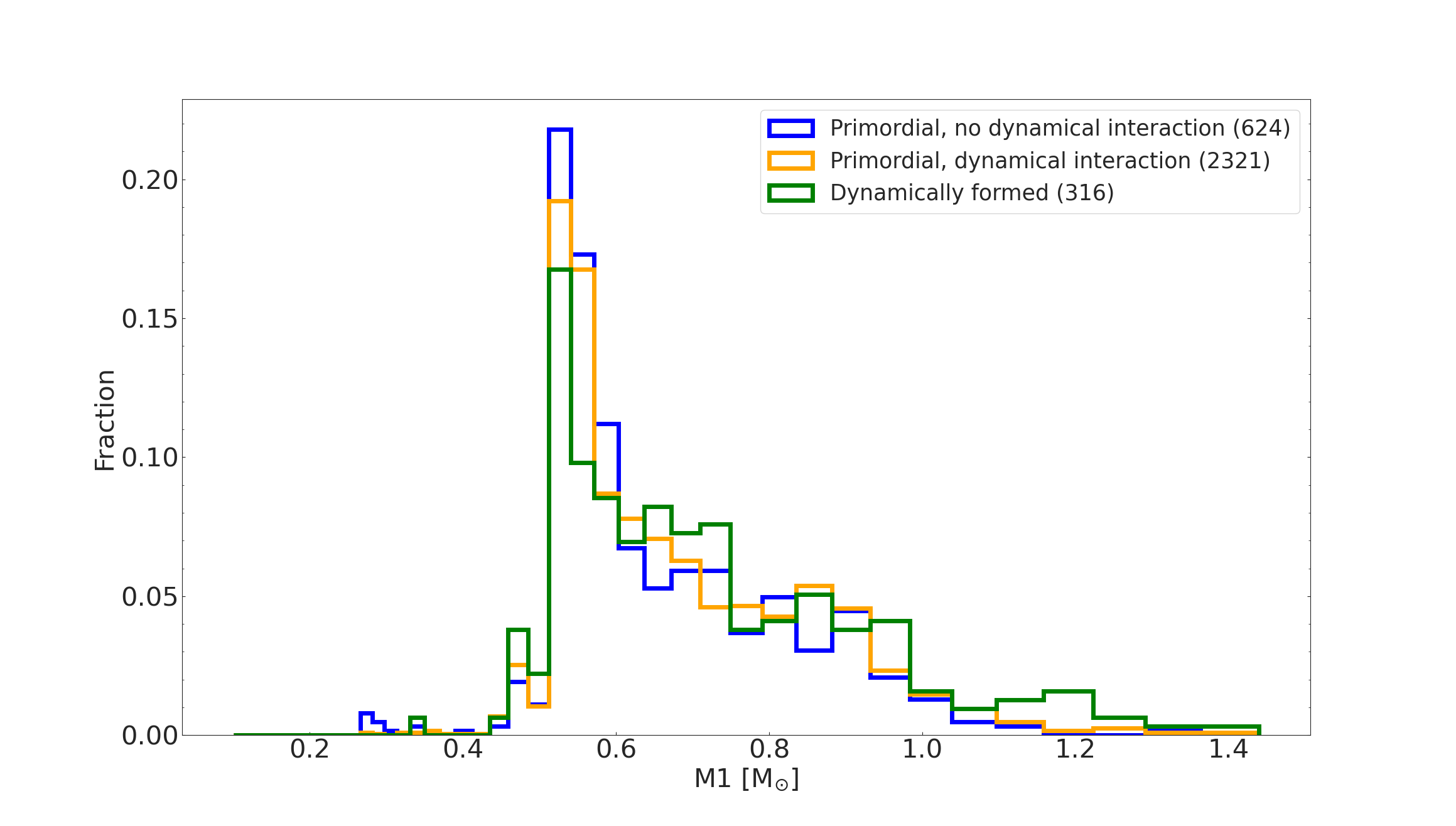}
      \caption{}
    \end{subfigure}%
    \begin{subfigure}[b]{0.5\textwidth}
      \centering
      \includegraphics[width=.95\linewidth]{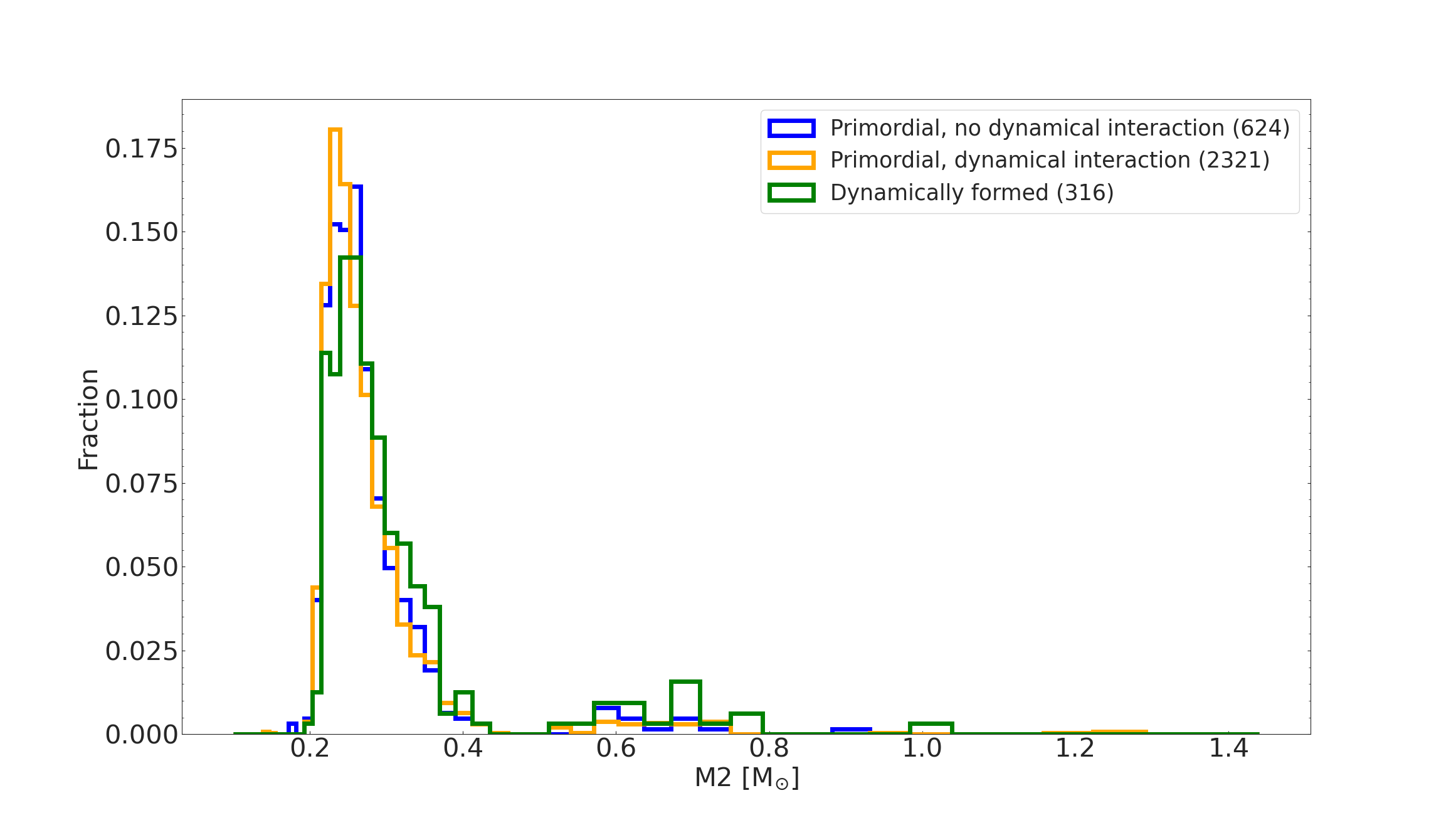}
      \caption{}
    \end{subfigure}
    \begin{subfigure}[b]{0.5\textwidth}
      \centering
      \includegraphics[width=.95\linewidth]{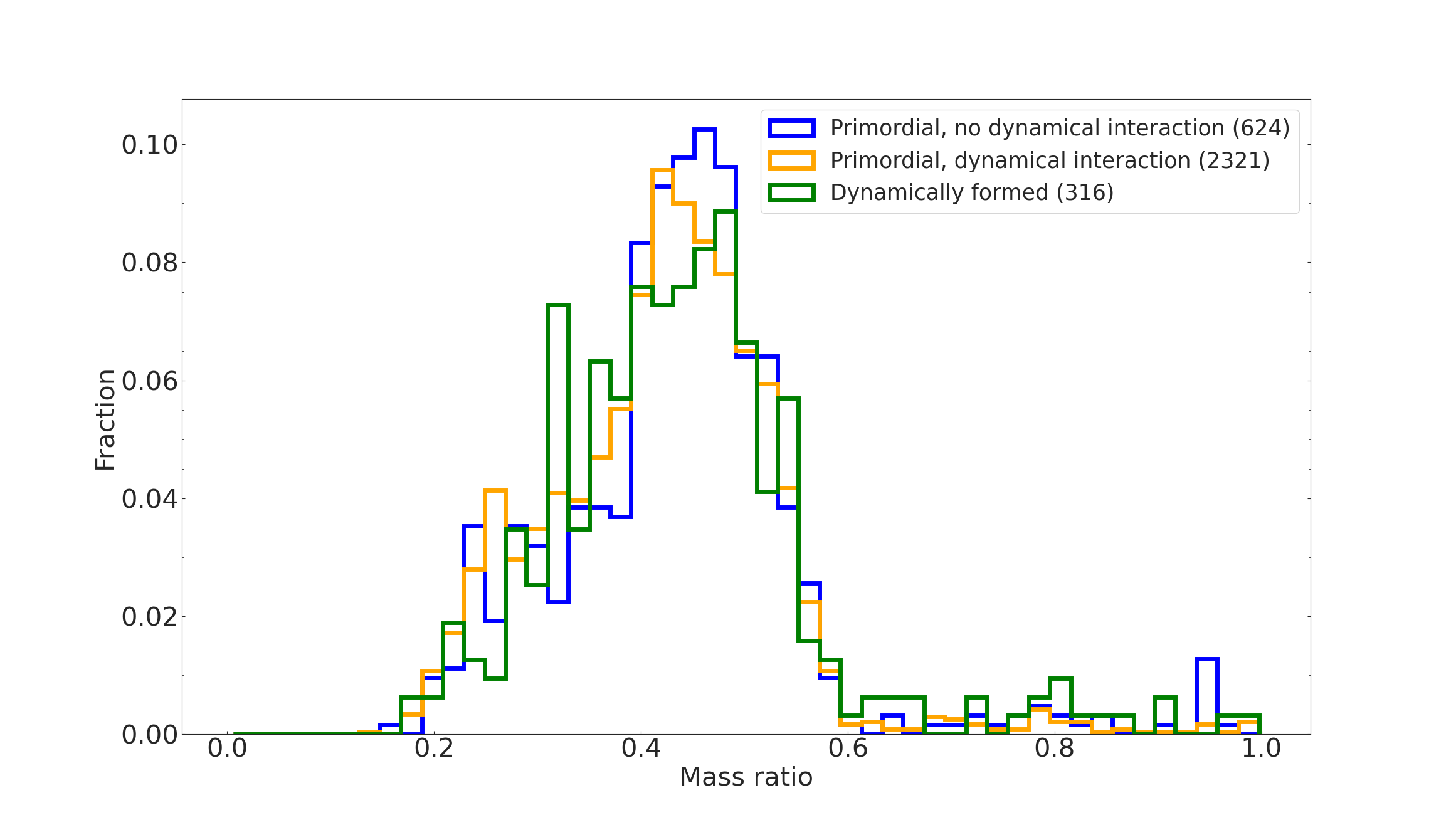}
      \caption{}
    \end{subfigure}%
    \caption{Histograms of the resolvable binary property distributions of the three formation channels: primordial binaries that were not involved in any dynamical interaction (blue), primordial binaries that were involved in dynamical interactions (orange) and dynamically formed binaries (green). Panel a) shows the semi-major axis, panel b) shows the eccentricity in log scale to highlight outliers, panels c) and d)
shows the primary and secondary mass respectively and panel e) shows the mass ratio $\left(\frac{M2}{M1}\right)$.}
    \label{fig:binaryProperties_resolvable}
\end{figure*}

We have split the set of resolvable DWDs into three formation channels: primordial binaries that did not experience a dynamical interaction, primordial binaries that did experience a dynamical interaction and dynamically formed binaries. In Fig.\ref{fig:binaryProperties_resolvable} we plot the parameters of these binaries. From panel a) we can see that the difference in semi-major axis is very small between the three channels, this is expected since the SNR is highly dependent on semi-major axis and limiting in SNR will also lead to a limit in semi-major axis. However, we see some differences at large semi-major axis; binaries that were involved in dynamical interactions have slightly higher separations compared to the ones that are from pure stellar/binary evolution. The reason for this is found in panel b) where we can see that the eccentricity of some binaries is non-zero. A tight DWD that formed in isolation can not be eccentric due to circularisation during the common-envelope phase, however a tight DWD can have its eccentricity increased in dynamical interactions such as binary-single or binary-binary interactions. It is also possible to form new binaries in dynamical interactions through exchanges, these binaries can have larger eccentricity than the primordial ones. This means that eccentric binaries formed through dynamical interactions can remain detectable (with SNR > 10) even at larger separations compared to circular binaries. Since almost all resolvable binaries are circular, the three peaks at $e=0$ are almost identical and overlap.

The panels c) and d) do not statistically relevant differences between the channels. In panel e) we see small differences in the mass ratio between the channels but this is most likely due to statistical fluctuations caused by a small dataset. 

In Table \ref{tab:originAndNumbers} we show the number of dynamically formed and primordial resolvable eccentric binaries. For the dynamically formed binaries we also show the number of DWDs that had at least one additional dynamical interaction after their formation, the number of binaries which were formed before both components had evolved into WDs, and the number of binaries that were directly dynamically formed with two WDs. We can see that the number of dynamically formed (24) and primordial binaries (21) are very similar so it seems like no formation channel is more dominant over the other. In addition to this we can see that most dynamically formed DWDs were directly formed with two WDs and did not experience any further interactions after their formation. This may be due to getting kicked out from the center of the cluster after the formation of the binary or due to the very small semi-major axis and low merger time where the binary merges relatively shortly after it forms.

\begin{table*}
\caption{Number of eccentric resolvable DWDs for two different formation channels. The third channel previously discussed (primordial without any dynamical interactions) cannot form these eccentric tight binaries and is thus not included here. The first column shows how the binary was formed, either primordial or dynamically formed. The second column shows the total number, the third column shows the number of binaries that were involved in interactions after both binary components had evolved into WDs and thus the DWD was formed. The fourth column shows the number of binaries that were dynamically formed before both components had evolved into WDs and the fifth column shows the number of DWDs that were directly formed in dynamical interactions.}
\label{tab:originAndNumbers}    
\centering                                      
\begin{tabular}{c c c c c}          
\hline\hline                        
Origin & Total \# & Had interaction after DWD formation & Not formed as DWD & Formed as DWD \\
\hline                                 
    Dynamical    & 24    & 11 & 5 & 19\\
    Primordial   & 21    & 21 & - & - \\
\hline                                             
\end{tabular}
\end{table*}

\begin{figure}[ht]
    \centering
    \includegraphics[width=0.5\textwidth]{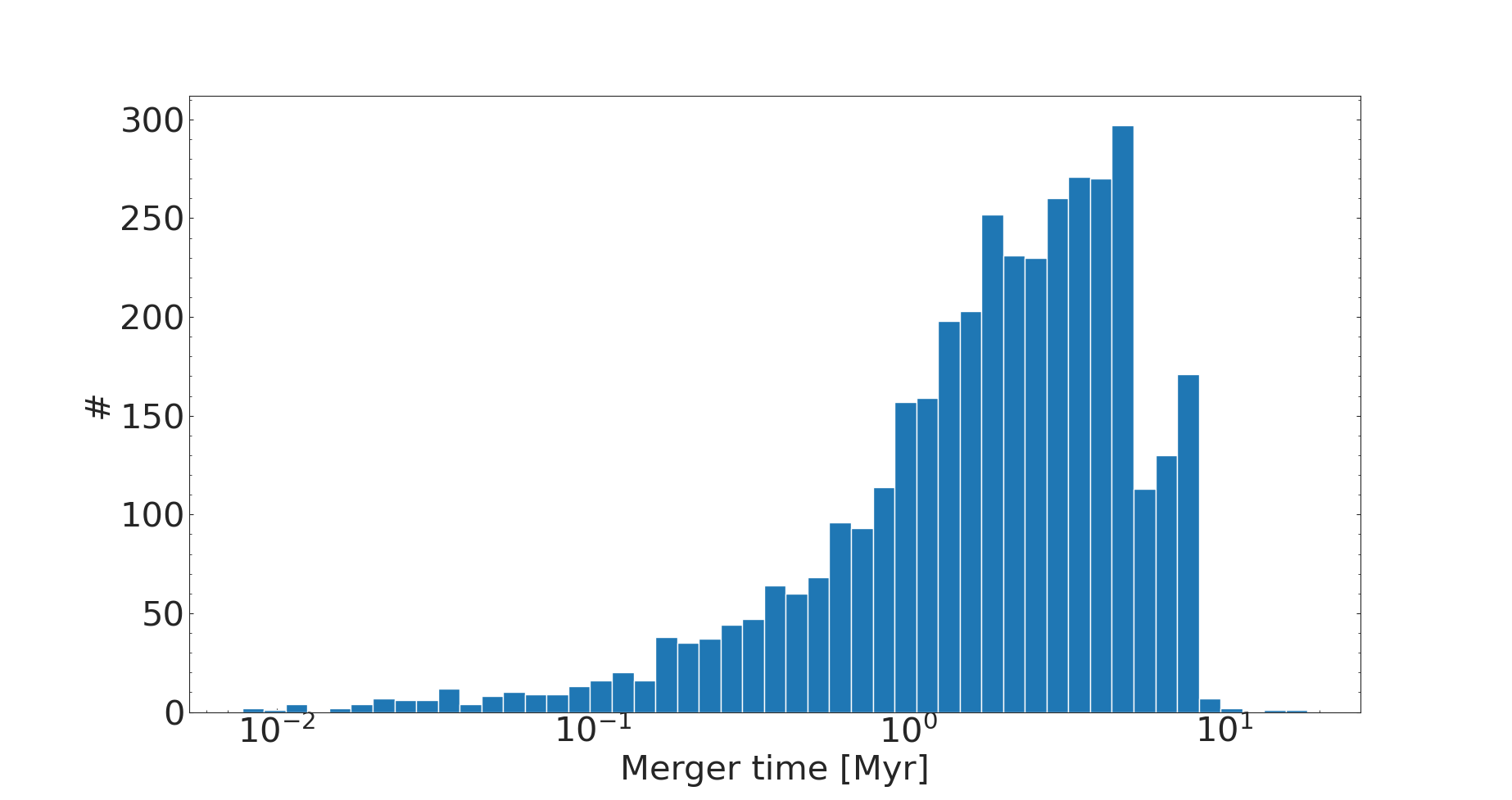}\\ 
    \caption{Distribution of merger time for the resolvable DWDs in our dataset.}
    \label{fig:mergerTime}
\end{figure}

The merger time of tight DWDs binaries is important since it will affect the probability to observe them. A low merger time gives a smaller window of possible detection. Fig.\ref{fig:mergerTime} shows the distribution of merger times for these binaries. We can see that they are all relatively short lived which is to be expected due the binaries being very tight. The longest merger time is 18 Myrs, with a large drop-off at $\sim8$ Myrs, and the shortest 7000 years. This reduces the chance of a successful detection since these binaries might merge before they are able to be detected.

\begin{figure*}[ht]
    \centering
    \includegraphics[width=\textwidth]{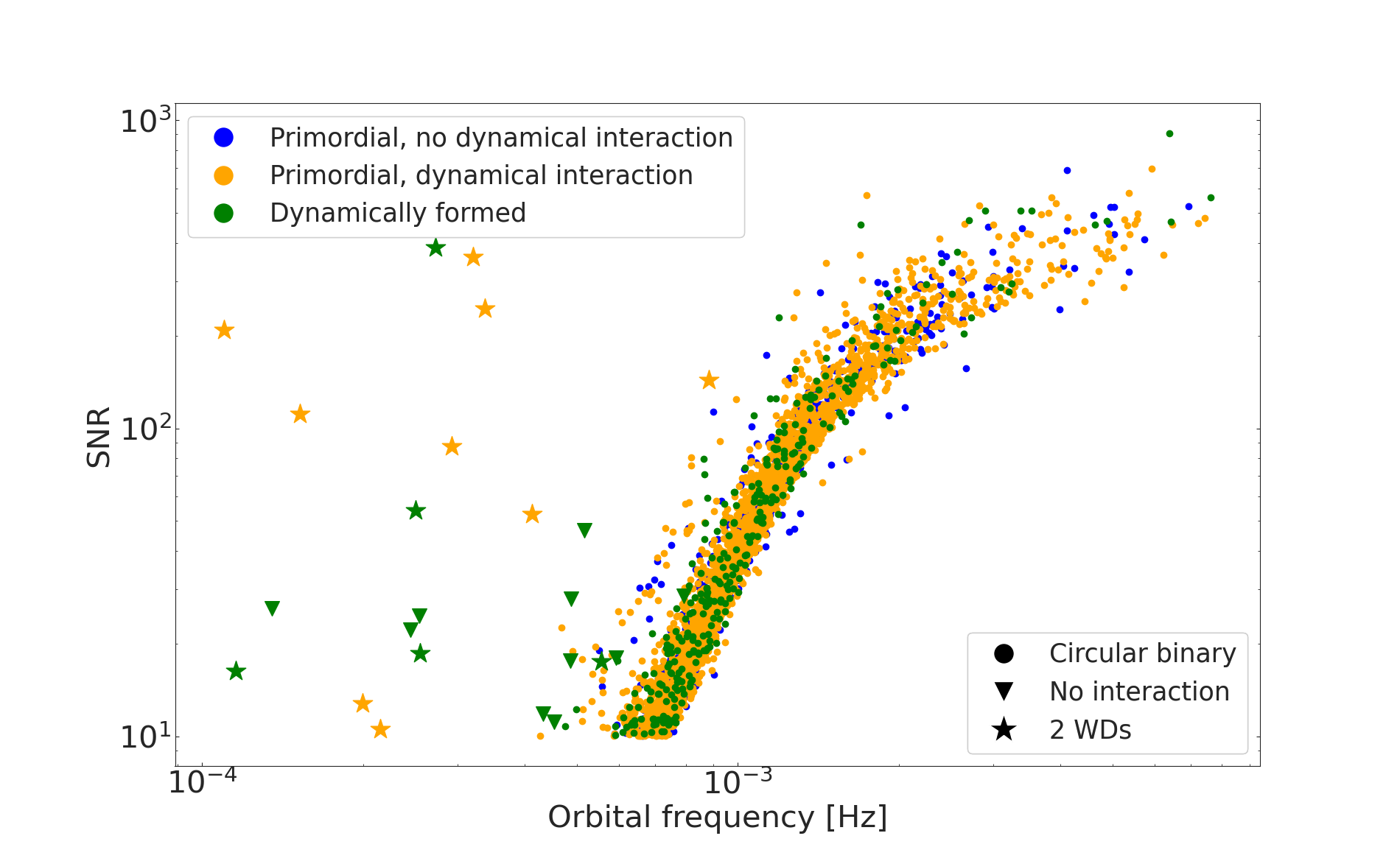}\\ 
    \caption{Orbital frequency plotted against signal-to-noise ratio for our resolvable binaries at a distance of 2 kpc and 5 years observation time. The different colors are explained in Fig.\ref{fig:binaryProperties_resolvable}. Circles represent circular binaries, triangles represent eccentric binaries that were not involved in any dynamical interactions after the formation of the DWD. Stars represent eccentric binaries that were involved in dynamical interactions after both components had evolved into WDs. The blue points follow the same curve as the orange points but are hidden due to the large number of points.}
    \label{fig:freq_snr_intInfo}
\end{figure*}

We plot the binary orbital frequency against the SNR in Fig.\ref{fig:freq_snr_intInfo} for every binary in these clusters between 9 and 14 Gyrs. We see that DWDs follow a distinct trend, which comes from their predictable evolutionary path governed by well-understood processes such as gravitational wave emission and orbital decay. The reason behind this is the dependence of the characteristic strain on the GW frequency of the binary. If considering binaries at different distances from the observer the trend might be weakened. This predictability is very useful in some cases but causes DWD from clusters and from the field to be very similar and difficult to distinguish from each other. The majority of DWDs are found along this trend, these DWDs are, in almost all cases circular and, due to our limit on SNR, very tight. They mostly have masses ranging from 0.2 to 1.4 M$_{\odot}$ with varying combinations of the mass ratio.

To the left of this big group we see some outliers, as either a triangle or star. 
These binaries are eccentric with eccentricities up to 0.99. This eccentricity increases the SNR due to a close pericenter distance which means that eccentric binaries may have lower orbital frequencies than circular binaries while still emitting GWs as SNR $>$ 10. The important thing to note here is that all of these eccentric binaries were involved in dynamical interactions and as such cannot be formed in the field. This allows us to distinguish between binaries from clusters and from the field. However, since these binaries are eccentric, they emit GWs in different frequencies due to the different harmonics. Taking this into account may bring them closer to the trend. We will look into this in the next paper.

To see more clearly how dynamical encounters are important for the formation of these DWD systems we mark the points with different markers depending on if they were involved in any dynamical interaction and if so, in which stage of their binary evolution they were involved in the interaction. Triangle means no interaction and a star means that both binary components were WDs. All of our tight and eccentric primordial binaries are involved in dynamical interactions after both components have evolved into WDs. This shows the importance of dynamical interactions after the formation of the DWD. There are some green triangles, these are dynamically formed binaries that directly formed with two WDs at a high eccentricity. After the formation they did not experience any more dynamical interactions. We can clearly see that, for primordial binaries, without dynamical interactions after both WDs are formed, we are not able to form these eccentric tight binaries. This is expected since even if a dynamical interaction increases the eccentricity of a MS binary, the common-envelope phase or phases would circularize it again. We do not differentiate between binary-single and binary-binary interactions here, we only take into consideration the stellar types of the binary in question when the interaction occurred.

We can see that there are multiple binaries that were involved in dynamical interactions and for some binaries this occurred after both WDs were formed but have low eccentricity and are thus not distinguishable from the trend. These binaries had very weak flybys where the binary properties were not significantly changed.

To summarize, for primordial binaries, the formation channel of tight and eccentric systems that are resolvable is as follows:
\begin{enumerate}
    \item The binary must form with a small semi-major axis or be hardened by dynamical interactions.
    \item There needs to be strong mass transfer during the evolution to tighten the binaries to the point where the GW signal from them will be resolvable. This mass transfer will most likely happen during two common-envelope phases.
    \item After both components have evolved into WDs, the binary needs to be involved in dynamical interactions which can increase the eccentricity.
\end{enumerate}

For dynamically formed binaries we have two formation channels:
\begin{itemize}
    \item Direct formation of two WDs, forming a binary with small semi-major axis and high eccentricity. In this scenario there is no need for additional dynamical interactions after the formation of the DWD since the binary is already tight and eccentric.
    \item Dynamical formation of a tight binary where both objects have not evolved into WDs. In this scenario we do require additional dynamical interactions after both stars have evolved into WDs. 
\end{itemize}


\section{Discussion \& Conclusions}
\label{sec:discussion}
As we have shown, it would be very difficult to distinguish strong GWs originating from circular DWDs in clusters and in the field. Instead, by focusing on eccentric tight binaries we can see differences. Since these binaries cannot be formed without dynamical interactions, a high density system is generally required to increase the eccentricity. This means that if we would find a signal that can be inferred to come from an eccentric DWD we can conclude that this signal should originate from a cluster. However, we might be interested in knowing how probable finding such a binary is. As seen in Fig.\ref{fig:mergerTime}, the lifetime of these binaries are relatively low and thus the window to observe them is small. In Table \ref{tab:probabilityEccBin} we show the number of eccentric resolvable binaries for different snapshots around 12 Gyrs; within 50 Myrs, within 100 Myrs and within 500 Myrs. We also show the number for all times after 9 Gyrs. The second column shows the total number of binaries, the third column shows the expected number of eccentric resolvable DWDs by cluster, obtained with the fraction of resolved binaries divided by the total number of binaries, and the fourth column shows the expected number by unit mass, obtained by taking the number of resolved binaries divided by the total mass of the clusters.

\begin{table*}
\caption{Probabilities to observe a resolvable high eccentricity DWD. The first column shows the snapshot in time, the second column shows the number of binaries in this snapshot from all models, the third column shows the expected number by cluster, given by the number of resolvable eccentric DWDs divided by the number of models and the fourth columns shows the expected number by unit mass, given by the number of resolvable eccentric binaries divided by the total mass of the models used.}       
\label{tab:probabilityEccBin}    
\centering                                      
\begin{tabular}{c c c c c}          
\hline\hline                        
Time [Myr] & \# binaries & Expected number by cluster & Expected number by unit mass [$1/M_{\odot}$] \\    
\hline                                 
    11950 - 12050   & 2    & $1.3 \times 10^{-2}$ &  $5.6\times 10^{-8}$\\
    11900 - 12100   & 3    & $1.9 \times 10^{-2}$ &  $8.5\times 10^{-8}$\\
    11500 - 12500   & 9    & $5.8 \times 10^{-2}$ &  $2.5\times 10^{-7}$\\   
    $\geq 9000$ Gyr & 43   & $2.3 \times 10^{-1}$ &  $1.2\times 10^{-6}$\\
\hline                                             
\end{tabular}
\end{table*}

 When looking at a smaller snapshot (50, 100 or 500 Myrs) we never see more than 1 tight and eccentric DWDs in one cluster at one time. This is due to short lifetime. Therefore, it's rare to observe an eccentric and tight DWD inside a cluster due to the relatively short window of opportunity. Even though the probability to observe an eccentric resolvable DWD is fairly small, this is for one cluster and when taking many clusters into consideration we expect to find a few of these binaries in the Milky Way (MW). Assuming a total GC mass of $10^8$ M$_{\odot}$ in the MW \citep{harris1996}, this gives an upper limit of $~5-6$ up to $\sim25$ depending on the size of the snapshot. Although this needs to be taken as a upper limit, we put all our clusters at 2 kpc since the purpose of this paper is to see if and what kind of DWDs in GCs that we could observe using GWs rather than giving a more precise estimate for the number. We will look into more details in the next paper. It was shown in \cite{hypki2025} that the models we use in this study are good approximations for MW GCs, although, we are not doing any direct comparisons to real clusters. The purpose of this paper is to explore the possibility to observe such a binary and the formation channels required to form it. Additionally, we put all binaries at 2 kpc and did not account for different distances to the observer and thus our results are an upper limit to the probability.

\begin{figure*}[ht]
    \centering
      \includegraphics[width=0.5\textwidth]{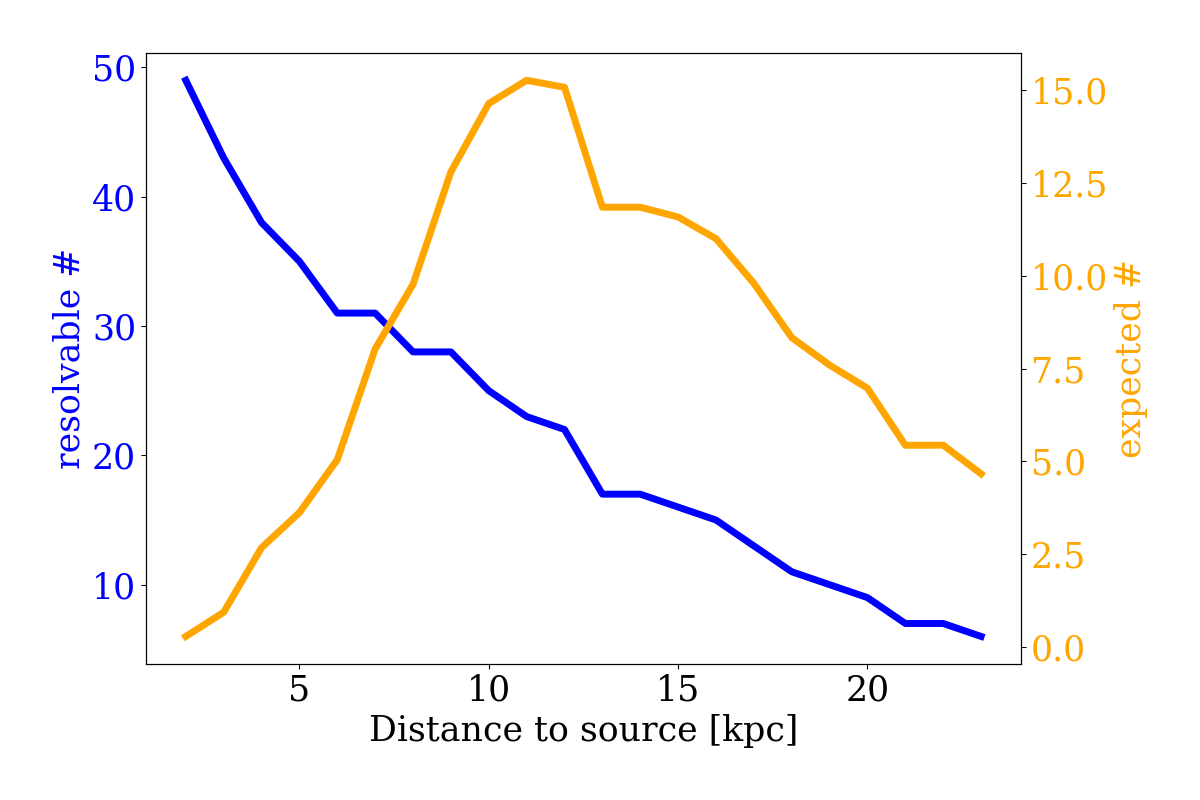}\\ 
     \caption{Number of resolvable eccentric binaries (blue) and upper limit of expected number of resolvable eccentric binaries in the MW (orange) against the distance to the source.}
    \label{fig:dist_resolv}
\end{figure*}

To explore how distances affects the probability to observe a tight and eccentric binary we find the number of resolvable binaries at different distances by calculating the SNR of all binaries from all models at different distances from the observer and assume that he MOCCA models are representative for the MW GC population. The binary properties and observation time is kept constant. This is shown in Fig.\ref{fig:dist_resolv} where the number of resolvable binaries are shown in blue and in orange we show a expected number of resolvable eccentric DWDs in the MW. To obtain this we multiply the expected number by unit mass with the total mass of GCs within a given distance. With this we can see that due to the increased amount of mass at larger distances the number, averaged by mass, of resolvable eccentric binaries increases up until 11 kpc and then starts to decrease due to the SNR being decreased by the increasing distance to the source. From this we can see that it will be important to consider all nearby GCs when searching for GW signals from eccentric DWDs, not only the very closest ones.

This would be a very significant discovery since it would give an independent measure of the distance to the cluster, which combined with other methods of estimating the distance can be used to increase the certainty in the estimate. Distance estimates from these eccentric and tight DWDs will be discussed in a future paper. It could also give an indication about in which type of cluster we would expect to find this kind of binary. We tried to find any correlation between initial cluster properties and number of resolvable eccentric DWDs but were not able to find such a relation. This issue will be further discussed in the next paper.

We also need to to consider the fact that clusters are continuously losing members as escapers and this could be a way for clusters to populate the field with eccentric binaries. We investigated this by selecting the escaping DWDs from all of these models and checking their SNR within at 12 Gyrs. From this we could not find any escaping eccentric tight DWDs that are resolvable at 12 Gyrs. Most binaries escape at an earlier time and if they are tight enough to be resolvable while having high eccentricity, they would merge before the reach the age of 12 Gyrs. However, this is something that needs to be kept in mind when discussing since there is a small but non-zero probability that a binary like this would be kicked out from the cluster at the right time where it can be resolvable from the field. Although the probability for this is very low and we do not see this as a major problem for our conclusions.

As a final test we took a few long living models, similar to OCs, with only 1 stellar population and checked whether they host eccentric resolvable DWDs. Using the same estimates for observation period (5 years) and SNR threshold (SNR > 10), with distance set to 0.1 kpc since OCs are, in general, closer to us than GCs, we only find 1 eccentric and tight binary from 60 models in total. This binary then merged in 4 Myrs, making the window to observe this very small. This happened at around $t = 7300$ Myrs which is considered a very old age for an open cluster. Due to this we do not expect OCs to be a realistic formation region for eccentric and observable DWDs but this conclusion needs further confirmation which is outside the scope of this paper.

A consequence of MSP is that the first stellar population with low concentration drastically increases the number of escapers early in the evolution of the cluster, however, this is not a concern since at those early stages there are few DWDs formed and the ones that have formed would not have much time to get their eccentricity increased due to dynamical interactions. In addition to this, if a tight and eccentric binary escapes early from the cluster it would merge long before 12 Gyrs.

In the introduction we mentioned that eccentric and thigh DWDs could form from triple systems and while it is possible, it is unlikely. \mocca does not currently support triple systems, thus we are not able to do any direct analysis of this from our data. \cite{rajamuthukumar2025} gives an estimate of $10^4$ DWDs (from both triple and binary channels) in the MW and approximately 57\% of these binaries retain their tertiary star when entering the LISA frequency range. However, they also mention that it is expected that these tertiary stars are too distant to change the inner binary in a way that would be seen in their GW signal. In addition to this, triples formed in the field are usually planar, making the Kozai mechanism inefficient.

Our main conclusions can be summarized as follows:
\begin{itemize}
    \item Eccentric and tight primordial binaries can only be formed in a formation channel that is unique to dense environments which makes it possible to distinguish them from binaries in the field. This formation channel requires the binary to form with a small semi-major axis, experience strong mass transfer and two common-envelope phases and, after the formation of both WDs, dynamical interactions that increases the eccentricity.
    \item Binaries can directly form with two WDs, small semi-major axis and high eccentricity in either binary-binary, binary-single or three-body interactions. However, this typically requires dense environments.
    \item The resolvable eccentric DWDs are short-lived which decreases the window of opportunity of observing them. However, due to the large number of DWDs, there is still a good probability to find this kind of binary.
    \item The probability to find an eccentric and tight binary for one cluster at around 12 Gyrs is small, however, when considering many different clusters at varying distances from us, we should expect to observe an upper limit of 10-15 such binaries, see Fig.\ref{fig:dist_resolv}.
    \item Escaping binaries are unlikely to populate the field with eccentric and tight binaries since the lifetime is relatively short. The probability that the binary is kicked out at the correct time after the eccentric and tight DWD is formed and is then able to be observed and distinguished from the other field binaries is very low.
    \item Due to the shorter lifespans and lower densities of OCs, from our data, it is more likely to find a tight and eccentric DWD inside GCs than in OCs.
\end{itemize}

Finding such a binary would be a big step forward for both GW physics but also our understanding of GCs and distance measurements in the MW. The GWs from these binaries can be used as an independent estimate of the distance to their host cluster. In a future paper we will explore this idea using data from models similar to the MW GCs of today and analyzing the GW signal from them.


\section*{Acknowledgments}
LH, MG, AH, AA, GW were supported by the Polish National Science Center (NCN) through the grant 2021/41/B/ST9/01191.
AA acknowledges support for this paper from project No. 2021/43/P/ST9/03167 co-funded by the
Polish National Science Center (NCN) and the European Union Framework Programme for Research
and Innovation Horizon 2020 under the Marie Skłodowska-Curie grant agreement No.
945339. 
VVA acknowledges support from the Boya Postdoctoral Fellowship program of Peking University.
For the purpose of Open Access, the authors have applied for a CC-BY public copyright license to any Author Accepted Manuscript (AAM) version arising from this
submission


\section*{Data Availability}
Input and output data for the globular cluster simulations carried out in
this paper will be shared on request to the corresponding author. \footnote{\url{https://zenodo.org/records/10865904}}

\bibliographystyle{aa}
\bibliography{ref.bib}

\end{document}